\documentclass[aps,prl,twocolumn,superscriptaddress]{revtex4-2}

\usepackage{soul} 
\usepackage{graphicx}
\usepackage[utf8]{inputenc}
\usepackage{xcolor}
\usepackage{times}
\usepackage[caption=false]{subfig}
\usepackage{amsmath}
\usepackage{slashed}%
\usepackage{bbold}%
\usepackage{bm}
\usepackage{mathtools}
\usepackage{physics}
\usepackage{amsthm}
\usepackage{newunicodechar}
\usepackage{geometry}
\usepackage{soul}
\usepackage{simpler-wick}
\usepackage[compat=1.1.0]{tikz-feynman}
\usepackage[force]{feynmp-auto}
\usepackage{amssymb}
\usepackage{placeins}
\definecolor{javcolor}{RGB}{30, 144, 255}

\definecolor{nwred}{rgb}{1.0, 0.0, 1.0}

\definecolor{nwblue}{rgb}{0.0, 0.0, 1.0}

\definecolor{tgorange}{rgb}{1.0, 0.5, 0.0}

\definecolor{vbcolor}{rgb}{0.6, 0.0, 0.6}

\definecolor{gjggreen}{RGB}{0, 204, 0}

\newcommand{\Ag}{\mathrm{A}_{1\mathrm{g}}}
\newcommand{\Eg}{\mathrm{E}_{\mathrm{g}}}
\newcommand{\Tg}{\mathrm{T}_{2\mathrm{g}}}

\begin{document}


\title{Temperature-Dependent Raman Signatures of Local Symmetry Breaking in Vacancy Double Perovskite Cs$_2$SnI$_6$}

\author{Tomke E. Glier}
\email{tomke.glier@uni-hamburg.de}
\affiliation{Department of Physics, University of Colorado Boulder, Boulder, Colorado, 80309, USA}
\affiliation{Present address: Institute of Nanostructure and Solid State Physics, University of Hamburg, Hamburg, 22761, Germany}

\author{Nicholas J. Weadock}
\affiliation{Materials Science and Engineering Program, University of Colorado Boulder, Boulder, Colorado, 80303, USA}

\author{Julian A. Vigil}
\affiliation{Department of Chemistry, Stanford University, Stanford, California 94305, USA}

\author{Cole Shank}
\affiliation{Department of Physics, University of Colorado Boulder, Boulder, Colorado, 80309, USA}

\author{Xixi Qin}
\affiliation{Thomas Lord Department of Mechanical Engineering and Materials Science, Duke University, Durham, NC 27708, USA}
\affiliation{SLAC National Accelerator Laboratory}

\author{Gabriel J. Graf}
\affiliation{Thomas Lord Department of Mechanical Engineering and Materials Science, Duke University, Durham, NC 27708, USA}

\author{Hemamala I. Karunadasa}
\affiliation{Department of Chemistry, Stanford University, Stanford, California 94305, USA}
\affiliation{Stanford Institute for Materials and Energy Sciences, SLAC National Accelerator Laboratory, Menlo Park, CA 94025, USA}

\author{Volker Blum}
\affiliation{Thomas Lord Department of Mechanical Engineering and Materials Science, Duke University, Durham, NC 27708, USA}
\affiliation{Department of Chemistry, Duke University, Durham, NC 27708, USA}

\author{Michael F. Toney}
\affiliation{Materials Science and Engineering Program, Department of Chemical and Biological Engineering, Renewable and Sustainable Energy Institute, University of Colorado Boulder, Boulder, Colorado, 80309, USA}

\author{Dmitry Reznik}
\email{dmitry.reznik@colorado.edu}
\affiliation{Department of Physics, University of Colorado Boulder, Boulder, Colorado, 80309, USA}

\begin{abstract}Understanding and controlling the optical, thermal and mechanical properties of K$_2$PtCl$_6$-type double perovskites is essential for the design of lead-free, high-performance thermoelectric and optoelectronic materials. In this study, we investigate Cs$_2$SnI$_6$ using temperature-dependent near-resonance Raman scattering, complemented by density functional theory–based phonon calculations. The appearance of symmetry-forbidden phonon modes in the Raman response at cryogenic temperatures provides strong evidence for local symmetry breaking that does not develop into a global structural phase transition. Through detailed analysis of phonon intensities, energies, and spectral lineshapes, we characterize the temperature evolution of the symmetry-broken structural response of Cs$_2$SnI$_6$. Below 200~K, multi-phonon scattering is strongly enhanced, and symmetry-forbidden phonon modes emerge. The temperature dependence of phonon intensities follows a phenomenological Arrhenius behavior, revealing an activation energy of 15–20 meV. These results establish temperature-dependent Raman spectroscopy as a sensitive probe of local structural distortions and anharmonic structural dynamics in Cs$_2$SnI$_6$, providing a robust experimental basis for discussing the microscopic origins of symmetry breaking and electron–phonon coupling in halide double perovskites.
\end{abstract}

\maketitle

\subsection{Introduction}
Metal-halide perovskites (ABX$_3$, X = halide) have emerged as exceptional optoelectronic materials, noted for their long carrier lifetimes, high defect tolerance, and strong light–matter interactions. These properties are commonly associated with their soft and polar structures, in which electronic excitations are strongly coupled to atomic vibrations \cite{Puppin2020, Yamada2022, Miyata2017, Ghosh2020}. In lead-halide perovskites, the mechanically soft structure (particularly the corner-sharing metal–halide octahedral framework) can deform around photogenerated electrons and holes, enabling the formation of large polarons that extend over several unit cells \cite{Miyata2017, Munson2021, Guzelturk2021}. This strong electron–phonon coupling leads to effective structure-driven screening that shields charge carriers from scattering and trapping \cite{Jin2020}, and local polar fluctuations have been shown to be intrinsic to the metal-halide perovskite lattice, independent of whether the A-site cation is organic or inorganic \cite{Miyata2017, Yaffe2017, Reuveni2023}. Raman spectroscopy has played a key role in establishing this picture, as low-frequency Raman modes in lead-halide perovskites reveal pronounced polar fluctuations of the inorganic substructure that intensify at elevated temperatures and correlate with polaronic behavior \cite{Yaffe2017, Reuveni2023, Miyata2017_scienceAdv}. 

Cesium tin(IV) iodide (Cs$_2$SnI$_6$) is a lead-free double perovskite (A$_2$BB$^\prime$X$_6$ \cite{Wolf2021}) that has attracted attention for optoelectronics and energy-conversion applications, including photovoltaics and photodetectors \cite{Lee2014, Ullah2021, Chauhan2025, Kumar2024}. Cs$_2$SnI$_6$ is a native n-type semiconductor with a direct optical band gap of $\sim$ 1.3~eV at room temperature \cite{Lee2014, Maughan2016, Luo2020, Vigil2024}. In contrast to the tin(II) single-perovskite analogue CsSnI$_3$, Cs$_2$SnI$_6$ adopts the K$_2$PtCl$_6$-type perovskite structure \cite{Werker1939}, in which half of the single-perovskite B-site positions are vacant. This structural motif leads to spatially isolated [SnI$_6$]$^{2-}$ octahedra, which strongly influences the thermodynamic stability, electronic behavior, and structural dynamics, while maintaining a globally cubic ($Fm\bar{3}m$) average structure \cite{Arunkumar2018, Maughan2019}.

Unlike their ABX$_3$ counterparts, which typically undergo symmetry-lowering structural transitions below room temperature, many K$_2$PtCl$_6$-type double perovskites such as Cs$_2$SnI$_6$ retain this cubic structure down to cryogenic temperatures \cite{Maughan2016}. This structural robustness is accompanied by exceptionally low thermal conductivity, tunable band gaps within the optical spectral range, and enhanced air and moisture stability \cite{Maughan2019, Karim2019, Maughan2016, Lee2014, Vigil2024, Cai2017}. The characteristic structural dynamics are governed by a twofold rattling mechanism of Cs$^+$ ions and [SnI$_6$]$^{2-}$ octahedra and pronounced electron–phonon coupling \cite{Jong2022, Buhmla2022, Yamada2022, Wright2016, Ghorui2024}. In Cs$_2$SnI$_6$, these rattling modes contribute to strong lattice anharmonicity and enhanced phonon–phonon scattering \cite{Maughan2018, Ghosh2023}.
Despite growing theoretical interest in electron–phonon coupling and polaron-related effects in vacancy double perovskites, experimental access to the associated structural distortions remains challenging. Diffraction-based techniques primarily probe average crystal structures and are inherently insensitive to short-range static or dynamic local distortions \cite{Weadock2023, Beecher2016, Liu2019}. As a result, experimental signatures of local symmetry breaking may remain hidden even when such distortions play an important role in the structural and electronic response.

Raman spectroscopy offers a powerful and complementary approach to this problem. As a symmetry-sensitive probe governed by well-defined selection rules, Raman scattering provides direct access to atomic vibrations and their coupling to electronic excitations. In particular, the appearance of symmetry-forbidden phonon modes or enhanced multi-phonon scattering can serve as sensitive indicators of local structural distortions and strong electron–phonon interactions \cite{Saunders2025, Saunders2022}.

In this work, we employ temperature-dependent, near-resonant Raman spectroscopy to investigate the structural dynamics of single-crystalline Cs$_2$SnI$_6$. We observe clear signatures of local symmetry reduction in the vibrational response, manifested by the activation of phonon modes that are symmetry-forbidden in the cubic structure and by a strong enhancement of multi-phonon scattering. By systematically tracking the energies, linewidths, and intensities of these modes as a function of temperature, we characterize the temperature-dependent behavior of the symmetry-broken structural response.

\subsection{Materials and Methods}

\paragraph{Raman Measurements.} We utilize a 671~nm (1.85 eV) CW laser source at a laser power of 2~mW and a spot size of the order of 100~µm. This laser energy is above the direct band gap of Cs$_2$SnI$_6$ of $\sim$ 1.3~eV to take advantage of resonance-Raman effects. Due to the ultralow thermal conductivity of these materials, detailed laser power studies have been conducted to exclude sample damage or extreme laser heating (see Fig. S2). The Raman instrument consists of a McPherson Scanning Monochromator (Model 207), equipped with a liquid-nitrogen cooled Pylon 100BRX (Teledyne) CCD detector. The Raman signal is collected in backscattering configuration. The elastic line is filtered by using three Bragg-Notch filters (Optigrate BNF-671-OD3-11M). Temperatures between 6~K and 295~K were obtained by a Janis closed-cycle exchange-gas cryostat system.
\paragraph{Preparation of Cs$_2$SnI$_6$ Crystals.} Millimeter-scale crystals of Cs$_2$SnI$_6$ were grown by slowly cooling (1-2~$^\circ$C~h$^{-1}$) a saturated solution of the precursors in dry $\gamma$-butyrolactone from 130~$^\circ$C, as reported previously \cite{Vigil2024}. The Cs$_2$SnI$_6$ crystals were isolated from the mother liquor and stored in the dark in a dry atmosphere; crystals were cleaved with a blade prior to Raman measurements, revealing a fresh (111) termination.
\paragraph{Phonon and Raman calculations.}
All density functional theory (DFT) calculations in this work were performed using the all-electron electronic structure code FHI-aims \cite{blum2009ab}  with numeric atom-centered orbital (NAO) basis sets. Two structures were considered for the phonon calculations: the experimental structure with space group $Fm\bar{3}m$ measured at 100~K, and a relaxed structure obtained by first applying a small displacement along one of the three degenerate imaginary $\Gamma$-point phonon modes of the 100 K experimental structure, followed by relaxation and symmetrization of the atomic positions while keeping the lattice constants fixed to the 100~K values. 
For the experimental structure, phonon calculations were performed using a 1×1×1 supercell of the conventional unit cell (36 atoms) with atomic displacements of 0.005 Å. The resulting force constants were folded back to the primitive cell (9 atoms) to obtain phonon energies and irreducible representations of $\Gamma$-point modes.
The relaxation was performed using the Perdew-Burke-Ernzerhof (PBE) exchange-correlation functional \cite{perdew1996generalized} with the Tkatchenko–Scheffler (TS) pairwise dispersion correction \cite{tkatchenko2009accurate} including the alkali modification introduced and benchmarked in \cite{Kim2022tso}. This level of theory has been shown to predict structures for metal-halide perovskites in close agreement with experiment in numerous past studies by our group \cite{Weadock2023tct,Biliroglu2026ush,Chakraborty2026lsb}. FHI-aims' "intermediate" basis sets and other numerical defaults were employed and the k-grid used was 4×4×4. Structural relaxation was considered complete once the maximum residual force on every atomic coordinate was below 0.0001 eV/Å. The atomic positions were symmetrized and the structure was brought to a primitive cell in the new space group (18 atoms) using Pymatgen's spglib interface \textit{SpaceGroupAnalyzer}  \cite{ong_python_2013, togo_spglib_2024}.
Phonons were computed in a 3×3×2 supercell of the primitive unit cell (324 atoms total) on a 2×2×2 k-grid, and the force constants were folded back to the primitive cell. Displaced structures were generated using the phonopy code \cite{togo_first_2015}, and forces were calculated at the DFT-PBE+TS$\_$alkali level-of-theory with the “intermediate” basis sets. Born effective charges (BECs) for non-analytical term correction (NAC) were calculated in the primitive cell using the same level-of-theory, basis sets, numerical settings, and a 6×6×4 k-grid. The dynamical matrix was then constructed from the force-constant matrix and used to determine phonon frequencies and eigenvectors. Raman spectra for the relaxed structure were calculated using density-functional perturbation theory (DFPT) \cite{shang2018all}, by evaluating the finite-difference polarizability gradients in the primitive cell using the same basis sets, numerical settings, and a 6×6×4 k-grid.

\begin{figure*}[htp!]
\includegraphics[width=0.85\linewidth
		]{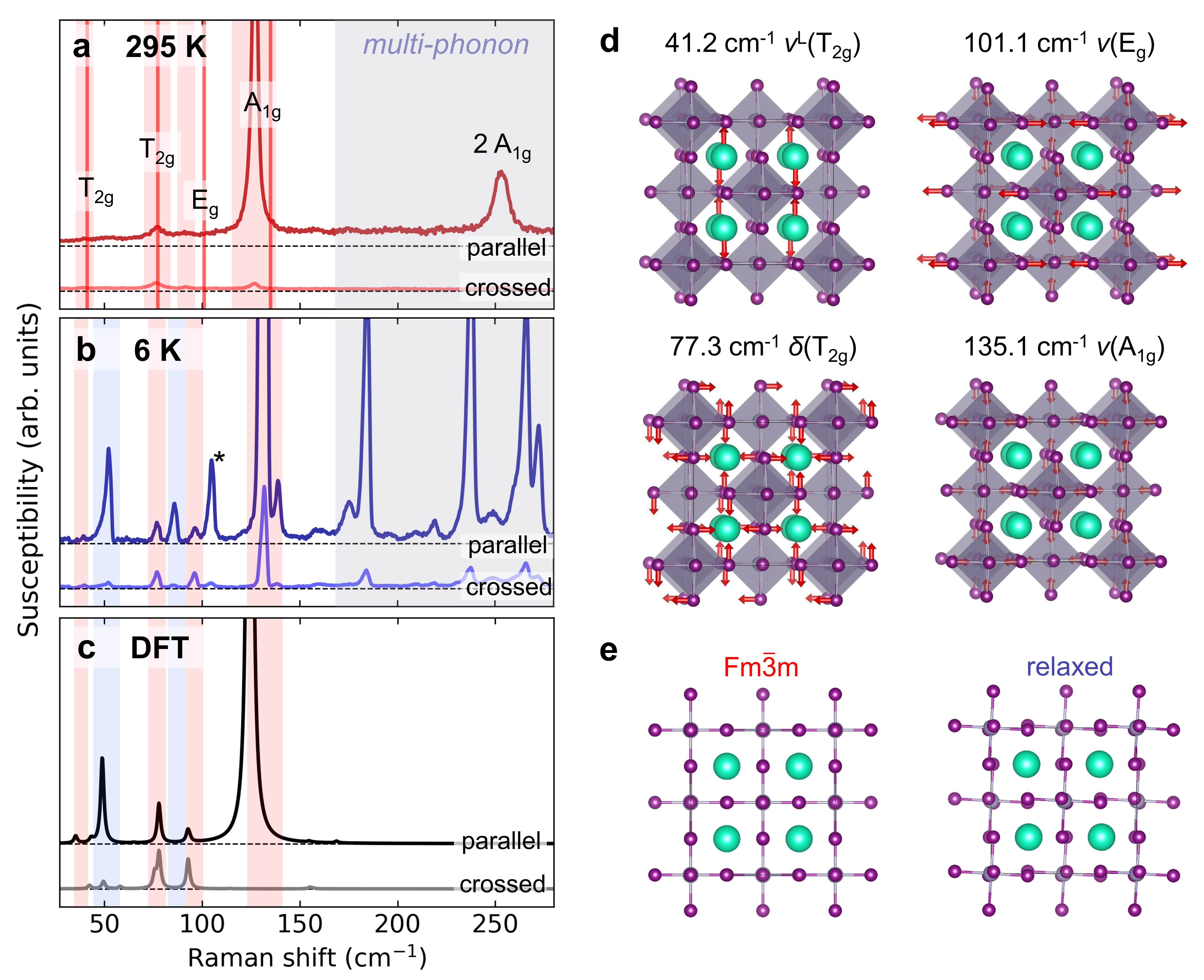}
	\caption{\label{fig:fig1} Cs$_2$SnI$_6$ Raman data. (a) Raman spectra at 295 K measured in $z(yy)\bar{z}$ (upper curve, $I_{\parallel}$) and $z(yx)\bar{z}$ (lower curve, $I_{\perp}$) polarization. Solid red vertical lines indicate the positions of the Raman-active phonons expected for the cubic $Fm\bar{3}m$ structure. Experimentally observed modes that can be assigned to these symmetry-allowed phonons are highlighted in red, and their corresponding irreducible representations are indicated. b) 6~K Raman spectra for $z(yy)\bar{z}$ (upper curve, $I_{\parallel}$) and $z(yx)\bar{z}$ (lower curve, $I_{\perp}$) polarization. Phonon modes that emerge at cryogenic temperatures compared to room temperature are highlighted in blue. The star denotes a feature previously discussed as a combination mode \cite{Folgueras2021, Kaur2022}. Dashed horizontal lines in (a) and (b) indicate zero Raman susceptibility. In the gray shaded region, multi-phonon features are observed. All Raman spectra are corrected for the temperature-dependent Bose function. c) DFT calculated Raman intensity for the relaxed structure obtained by relaxing and symmetrizing the atomic positions while fixing at the 100 K lattice constant, after initiating the symmetry-breaking relaxation by applying a small displacement along one of the three degenerate imaginary $\Gamma$-point phonon modes. Parallel and crossed scattering geometries have been calculated relative to the (111) facet matching the experimental conditions.
    d) Raman-active phonon modes and calculated Raman shifts corresponding to the experimental $Fm\bar{3}m$ crystal structure of Cs$_2$SnI$_6$. The $\Gamma$-point phonon energies and irreducible representations were obtained from finite-displacement DFT phonon calculations using a $1\times1\times1$ supercell of the conventional unit cell. e) Experimental $Fm\bar{3}m$ and relaxed structure for DFT. See Methods for details.} 
    \end{figure*}

\subsection{Results}
A$_2$SnI$_6$ perovskites consist of isolated SnI$_6$ octahedra and lack Sn–I–Sn bridges. As a result, Raman-active modes are dominated by internal octahedral vibrations (Sn–I stretching and bending), while low-frequency phonons are less prominent than in ASnI$_3$, where corner-sharing octahedra enable strong Sn–I–Sn vibrational coupling \cite{Jong2022, Kontos2017, Bhui2022}. Additionally, the high symmetry cubic structures typical of A$_2$BX$_6$ compounds impose stricter Raman selection rules, resulting in fewer observable Raman-active modes \cite{Rahim2020}.

Figure~\ref{fig:fig1} shows representative Stokes Raman spectra of Cs$_2$SnI$_6$ acquired at 295~K (Fig.~\ref{fig:fig1}a) and 6~K (Fig.~\ref{fig:fig1}b) in parallel and crossed polarization. All spectra are corrected for the thermal population factor (Bose correction). At 295~K, the Raman response is dominated by the four first-order modes expected for the cubic $Fm\bar{3}m$ structure and commonly reported for Cs$_2$SnI$_6$ \cite{Bhui2022, Kaur2022, Belessiotis2021}. Specifically, we observe the triply degenerate low-frequency $\nu^{L}(\Tg)$ mode near 39~cm$^{-1}$, the $\delta(\Tg)$ mode near 76~cm$^{-1}$, and the $\nu(\Eg)$ and $\nu(\Ag)$ stretching modes at approximately 96~cm$^{-1}$ and 132~cm$^{-1}$, respectively. In addition, a second-order feature assigned to $2\nu(\Ag)$ is visible near 260~cm$^{-1}$. The comparison of parallel and crossed polarization confirms the expected selection rules for scattering from the (111) facet (Table~\ref{tab:tab1} and Fig. S3) \cite{Cardona1982}.

\begin{table}
	\caption{Experimentally-observed Raman-active single-phonon modes of Cs$_2$SnI$_6$ along with theoretically-calculated modes for the experimental cubic $Fm\bar{3}m$ structure. Raman intensity ratios of crossed vs parallel polarization for scattering from the (111) crystal facet are denoted as $I_{\perp}/I_{\parallel}$ (see also Fig. S3). Experimental Raman shifts are given for the measurements taken at 6~K. To account for slight changes in experimental conditions between parallel ($\parallel$) and crossed ($\perp$) measurements, intensities are normalized to $I_{\perp}/I_{\parallel}$($\Eg$) = 1.0. \label{tab:tab1}}
    \begin{tabular}{p{0.5cm} | p{1.8cm} p{0.9cm} | p{1.8cm} p{1.6cm}} 
	\hline
        & \multicolumn{2}{c|}{\bf{Theory}} & \multicolumn{2}{c}{\bf{Experiment (6~K)}}\\
        \hline
	\bf{ } & Raman shift & $I_{\perp}/I_{\parallel}$ & Raman shift & $I_{\perp}/I_{\parallel}$\\
	\hline
	$\Tg$  & 41.2~cm$^{-1}$& $2/3$ & 39~cm$^{-1}$& $0.66 \pm 0.1$ \\ 
        $\Tg$  & 77.3~cm$^{-1}$& $2/3$ & 76~cm$^{-1}$& $0.71 \pm 0.08$ \\ 
        $\Eg$  & 101.1~cm$^{-1}$& $1$ & 96~cm$^{-1}$& 1.0 \\ 
        $\Ag$ & 135.1~cm$^{-1}$& $0$ & 132~cm$^{-1}$& $0.04 \pm 0.02$ \\ 
	\hline
\end{tabular}
\end{table}

Upon cooling, the Raman spectra evolve markedly: the overall phonon spectral weight increases and additional features emerge, yielding a substantially richer low-temperature spectrum (Fig.~\ref{fig:fig1}b). Besides the first-order modes already present at 295~K (red), additional peaks appear below $\sim$200~K, most prominently in the frequency range above $\sim$170~cm$^{-1}$. Here, enhanced by near-resonant excitation, a dense set of multi-phonon features is observed. Consistent with previous reports, we also resolve a feature near 105~cm$^{-1}$ that has been discussed as a combination mode \cite{Folgueras2021, Kaur2022}.
    
In addition to the enhanced multi-phonon response, three new low-energy features become clearly visible at low temperature (shaded blue): a mode near 50~cm$^{-1}$ with a strongly asymmetric lineshape and a mode near 86~cm$^{-1}$. Notably, these frequencies are close to the energies expected for infrared (IR)-active phonons of the ideal cubic $Fm\bar{3}m$ phase (Table~S2). The appearance of intensity at the frequencies of nominally IR-active modes in a Raman experiment indicates that the strict selection rules of the centrosymmetric high-symmetry structure are relaxed. This is commonly associated with local symmetry reduction, which can render modes that are symmetry-forbidden in the ideal structure observable in the Raman response \cite{Lee2022}.  The calculated vibration patterns shown in Fig.~S4 confirm that the relevant $T_{1u}$ and $T_{2u}$ modes correspond to symmetric vibrational motions of the SnI$_6$ octahedra that are Raman-inactive in ideal cubic $O_h$ symmetry. Any local deviation from this centrosymmetric environment, such as bond-length variations or distortions of the octahedral units, can lift inversion symmetry and lower the local point group symmetry (e.g., to $C_{2v}$ or $C_{3v}$)
. Under such reduced symmetry, the Raman selection rules are relaxed and formerly IR-active modes can acquire Raman activity, providing a natural explanation for the emergence of additional peaks near 50~cm$^{-1}$ and 86~cm$^{-1}$. At temperatures above 200~K, these additional low-energy features broaden and weaken, becoming increasingly difficult to distinguish from the background.

Fig.~\ref{fig:fig1}c presents the DFT-calculated Raman spectrum for the relaxed structure. This structure was obtained by relaxing the atomic positions at the fixed 100 K lattice constant after applying a small displacement along one of the three degenerate imaginary $\Gamma$-point phonon modes (see Methods and Fig.~\ref{fig:fig1}e). The resulting structure was found to exhibit $P2_1/c$ symmetry with a small numerical tolerance, and the atomic positions were refined accordingly, resulting in a negligible total energy change. The final structure and the corresponding Brillouin zone are shown in Fig.~S5. No remaining physically meaningful instabilities are observed in the phonon dispersion of this derivative structure (see Fig. S6). Thus, this procedure captures the dominant structural instability predicted by the phonon calculations and provides a representative local distortion pattern. The Raman tensors were evaluated for the parallel and crossed scattering geometries relative to the (111) facet, matching the experimental configurations. The calculated Raman spectrum was constructed by applying Lorentzian lineshapes with a phonon broadening of $\Gamma = 3$ cm$^{-1}$ to the DFT-derived mode frequencies and intensities. The relaxed structure reproduces the main experimental features observed at 6 K in good agreement, including the mode near 50 cm$^{-1}$. The calculated Raman intensities were rescaled by mode-dependent prefactors for comparison with experiment. Quantitative agreement of absolute intensities is not expected, as the DFT calculations do not include resonance Raman contributions present under the experimental excitation conditions. All calculated $\Gamma$-point mode frequencies and irreducible representations for the relaxed structure are presented in Table~S3. 

\begin{figure*}[htp!]
	\includegraphics[width=0.95\linewidth
		]{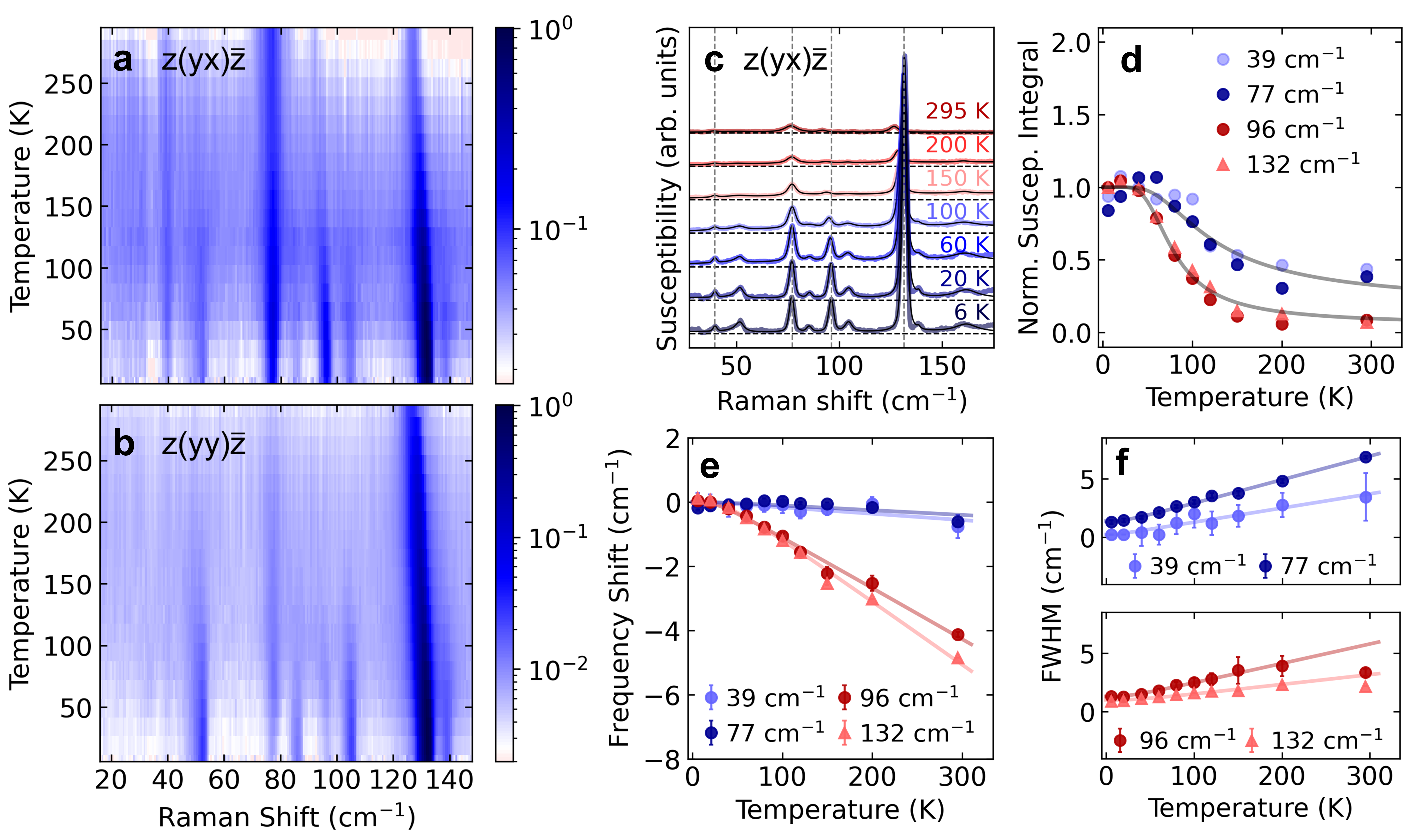}
	\caption{\label{fig:fig2} Raman susceptibility as a function of temperature. a) Interpolated 2D color plot for crossed polarization. b) Interpolated 2D color plot for parallel polarization. c) Exemplary Raman spectra between 10 and 140~cm$^{-1}$. Raman spectra are corrected for the temperature-dependent Bose function. The full data set for $z(yx)\bar{z}$ and $z(yy)\bar{z}$ polarization is shown in Fig. S1. d) Normalized phonon susceptibility integral extracted from the phonon fits. The curves are normalized to the respective susceptibility integral value at 6~K. e) Relative phonon frequency shift as a function of temperature for the four Raman-active phonon modes of the $Fm\bar{3}m$ structure (see Fig. 1). f) Phonon full width at half maximum of the Lorentzian width determined by a Voigt fit. The solid lines in e) and f) correspond to the Klemens model for a symmetric three-phonon anharmonic decay (see Eq. (2) and (3)) \cite{Klemens1966}. In all panels d) to f) the frequency, width and integral of the $\Ag$ mode are shown from parallel polarization (triangles), as this corresponds to the $\Ag$ character. Circles represent crossed polarization. Phonon energies indicated in the legends correspond to experimental values measured at 6~K.
 }
    \end{figure*}

Figure~\ref{fig:fig2} summarizes the temperature dependence of the four Raman-active single-phonon modes of the cubic $Fm\bar{3}m$ structure. The interpolated color plots in the low-frequency range in Fig.~\ref{fig:fig2}a and b illustrate the evolution of the Raman response in crossed and parallel polarization over the full temperature range. 
Selected Raman spectra in crossed polarization are shown in Fig.~\ref{fig:fig2}c. The Bose-corrected Raman spectra are fitted using Voigt profiles for the phonons and a tanh-function as background (see supporting information) to account for the nature of all Raman response functions that tend to 0 for $\omega \rightarrow 0$. For the Voigt profiles, the Gaussian contribution represents the instrumental response function, which is determined by measuring the elastic laser line. 

Figure~\ref{fig:fig2}d shows the temperature dependence of the integrated Raman susceptibility of the four Raman-active modes, normalized to their respective low-temperature values. Triangles for the $\nu (\Ag)$ mode represent results from parallel polarization corresponding to the $\Ag$ character. Circles are used for crossed polarization. For all modes, the integrated intensity decreases strongly with increasing temperature. The effect is particularly pronounced for the $\nu(\Ag)$ and $\nu(\Eg)$ modes, whose spectral weight is reduced by more than an order of magnitude between 6~K and 295~K, whereas the low-energy $\Tg$ modes exhibit a more moderate reduction.

We find that the temperature dependence of the Raman spectral weight can be described phenomenologically by an Arrhenius-type expression \cite{Soares2024, McCall2017, Rahane2024}:

\begin{equation} I_{Arr}(T) = \frac{I_0}{1+A \cdot \exp(-E_A/(k_B T))} \end{equation}

where $E_A$ represents an activation energy, $I_0$ is the phonon susceptibility integral at $T=0$ and $A$ is a rate parameter. The fits yield an activation energy of $E_A=20 \pm 3$~meV together with the rate parameter $A = 33 \pm 0.5$ for $\Ag$ and $E_\textrm{g}$ modes (96~cm$^{-1}$ and 132~cm$^{-1}$) and $A=6 \pm 1$ for the $\Tg$ modes (39~cm$^{-1}$ and 77~cm$^{-1}$).

The temperature-dependent phonon frequencies extracted from Voigt fits are displayed in Fig.~\ref{fig:fig2}e. The higher-energy stretching modes, $\nu(\Eg)$ and $\nu(\Ag)$, exhibit a pronounced softening with increasing temperature, amounting to a total shift of approximately 5~cm$^{-1}$ between 6~K and 295~K. In contrast, the two low-energy $\Tg$ modes show only minimal frequency changes over the same temperature range.

The linewidths (full width at half maximum, FWHM) obtained from the Lorentzian component of the Voigt profiles are plotted in Fig.~\ref{fig:fig2}f. For all modes, the linewidth increases with temperature, reflecting enhanced phonon scattering at elevated temperatures. To provide a quantitative description of the temperature evolution of phonon frequencies and linewidths, we compare the data to the Klemens model for anharmonic phonon decay, which describes the symmetric decay of an optical phonon into two acoustic phonons \cite{Klemens1966}. Within this framework, the temperature dependences of phonon frequency $\omega(T)$ and width $\Gamma(T)$ can be approximated by

\begin{equation}
\omega(T) = \omega_0 - \frac{a}{\exp(\frac{\hbar \omega_0}{2 k_B T})-1} \, ,
\end{equation}

\begin{equation}
\Gamma(T) = \Gamma_0 + \frac{b}{\exp(\frac{\hbar \omega_0}{2 k_B T})-1} \, ,
\end{equation}

where $\omega_0$ is the phonon frequency at $T = 0$~K, $\Gamma_0$ represents the FWHM at 0~K, and $a$ and $b$ correspond to phonon characteristic decay constants that contain the transition matrix element and the two-phonon density of states of the anharmonic decay \cite{Klemens1966, Cardona1984, Buchenau2020}. The model provides a phenomenological description of the observed softening and broadening of the $\nu(\Eg)$ and $\nu(\Ag)$ modes, while only very small decay constants are required to account for the nearly temperature-independent frequencies of the $\Tg$ modes. All extracted fit parameters are summarized in Table~S1.

Taken together, the analysis of the symmetry-allowed Raman modes reveals two key aspects of the structural dynamics: (i) a pronounced and mode-dependent anharmonic softening and broadening, and (ii) a strong thermally activated reduction of phonon spectral weight characterized by an energy scale of approximately 15–20 meV. Importantly, this energy scale emerges consistently from the temperature dependence of all cubic Raman-active modes, suggesting that the observed behavior is not mode-specific but reflects a common underlying process.

\begin{figure*}[htp!]
	\includegraphics[width=0.8\linewidth
		]{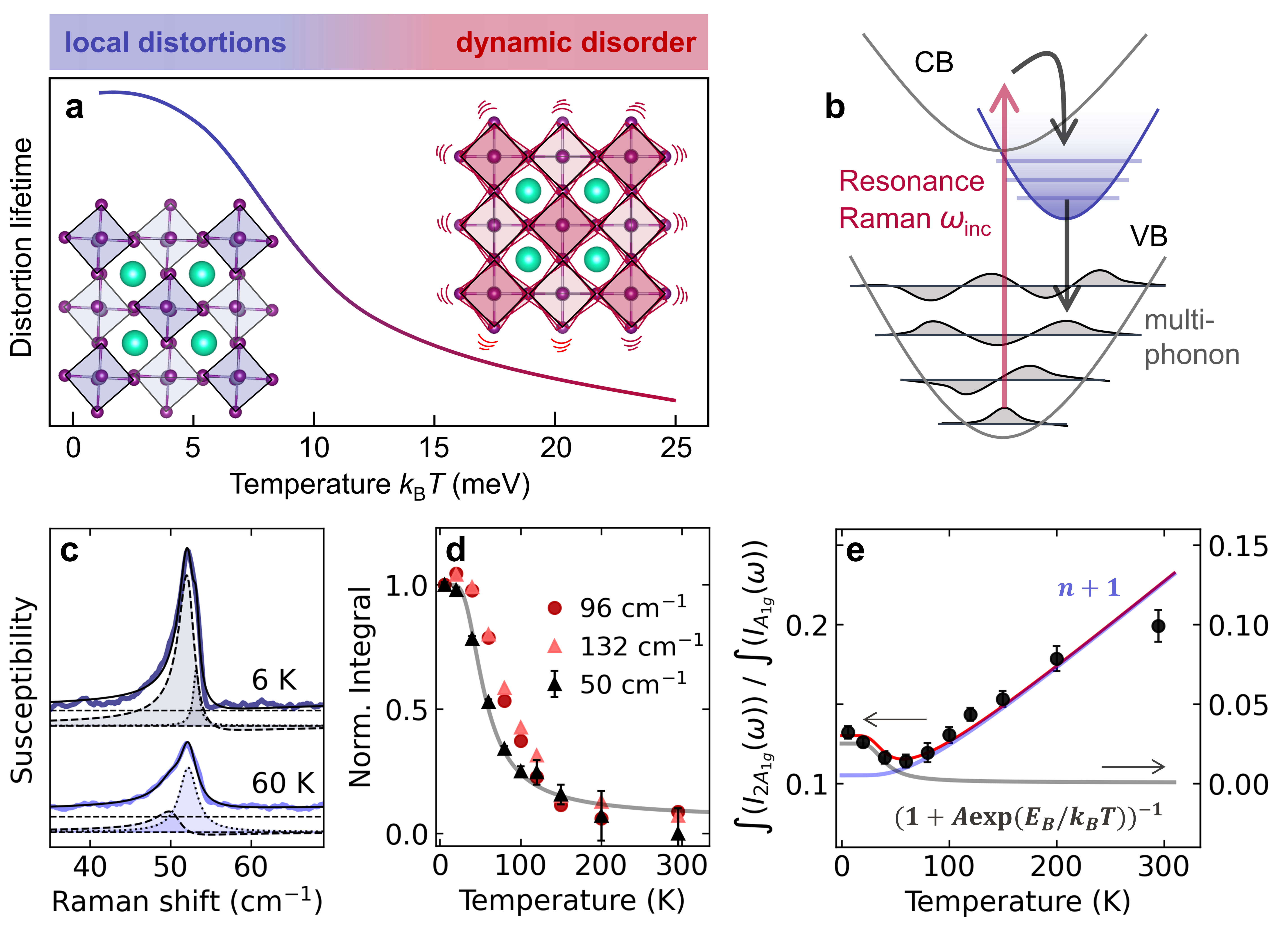}
	\caption{\label{fig:fig3} (a) Schematic illustration of the crossover from locally symmetry-broken / relaxed structural configurations at low temperatures to dynamically fluctuating disorder at elevated temperatures. The vertical axis represents the lifetime of locally distorted configurations as probed by Raman scattering. Upon increasing temperature, thermally activated structural fluctuations reduce the Raman visibility of local symmetry breaking. (b) Franck–Condon–like representation of resonant Raman scattering. Under near-resonant excitation ($\omega_{\mathrm{inc}}$), strong electron–phonon coupling leads to a displacement of the excited-state potential surface along a configurational coordinate. A self-trapped state at low temperatures, illustrated in blue and displaced along the configurational coordinate, leads to enhanced multi-phonon scattering.
    (c) IR-active phonon mode near 50~cm$^{-1}$ at 6 K and 60 K. The pronounced asymmetric lineshape is indicative of strong electron-phonon interactions. Fits to the data are shown as black solid lines. Lorentzian contribution is depicted as dotted line, Fano contribution is shown as dashed line. Most importantly, the 60 K data yield a Fano parameter of $q = -2.6$, whereas the 6 K data exhibit a weaker Fano strength with $q = -5.5$, consistent with reduced coupling to the electronic continuum at low temperatures. The curves are offset for better visibility. Dashed horizontal lines mark zero. (d) Normalized Raman susceptibility integral as a function of temperature for the $T_\textrm{1u}$ mode shown in (c) (black). For comparison, data for the $\Ag$ and $\Eg$ modes are also presented (red and light red, see also Fig. 2d). The gray solid line represents the Arrhenius equation with an activation energy $E_A = 15$~meV. (e) Integral ratio of two-phonon to single-phonon scattering intensity as a function of temperature for the $\Ag$ mode. At temperatures above 100~K, the data follow the expected Bose ratio of $n+1$ (blue solid line). At low temperatures, however, we observe an increased two-phonon Raman scattering intensity scaling with the localization of carriers in self-trapped states as fitted by the Arrhenius equation in Fig. 3d and shown as gray solid line. The sum of blue and gray solid line is shown as red solid line.
 }
    \end{figure*}

Figure~\ref{fig:fig3}a schematically illustrates the crossover from locally symmetry-broken structural configurations at low temperatures (blue) to dynamically fluctuating disorder at elevated temperatures (red). This crossover is reflected in the Arrhenius-like temperature dependence of the Raman intensities. In this picture, the intensity of symmetry-forbidden modes serves as an indicator of the lifetime of locally distorted configurations. Figure~\ref{fig:fig3}d summarizes the temperature dependence of the integrated Raman susceptibility of the mode near 50~cm$^{-1}$, together with the Raman-active $\nu(\Ag)$ and $\nu(\Eg)$ modes. All three exhibit a similar reduction in spectral weight with increasing temperature. The 50~cm$^{-1}$ mode follows an Arrhenius-type behavior (Eq.~(1)) with an activation energy of $15 \pm 3$~meV, comparable to the energy scale extracted for the cubic Raman-active modes (Fig.~\ref{fig:fig2}d). This common energy scale suggests that both symmetry-allowed and symmetry-forbidden features are influenced by a related thermally activated phonon process.
Consistent with this picture, cooling enhances the Raman visibility of locally symmetry-reduced configurations, while increasing temperature restores the effective cubic response through rapid structural fluctuations.

Strong electron–phonon coupling in the soft and polar perovskite structure can promote carrier localization at low temperatures. Such localized (self-trapped) configurations, illustrated schematically in Fig.~\ref{fig:fig3}b (blue potential), can influence the Raman response in two principal ways. First, changes in carrier localization can modify the effective polarizability response of the crystal structure, thereby affecting the overall Raman intensity \cite{McCall2017}. Second, resonant-Raman scattering experiments revealed a significant enhancement of multi-phonon scattering under resonance conditions. Previous studies have reported multi-phonon scattering up to the 12th order when exciting above the optical band gap of halide perovskites \cite{Iaru2021, Xu2022}. This phenomenon can be explained using a Franck-Condon-like picture (Fig.~\ref{fig:fig3}b), where strong electron-phonon interaction locally deforms the crystal structure near a photoexcited carrier, leading to its self-trapping. Consequently, the excited electronic state possesses different atomic coordinates than the ground state when excited above the band gap \cite{Perebeinos2001, Krueger2004, Andreasson2007}. As a result, we observe strong multi-phonon scattering as can be seen already in Fig. 1. The temperature dependence of the ratio between the two-phonon overtone and the corresponding single-phonon $\nu(\Ag)$ mode is shown in Fig.~\ref{fig:fig3}e. At temperatures above approximately 100~K, the intensity ratio follows the expected Bose population factor $(n+1)$. Upon further cooling, however, the ratio exceeds this expectation, indicating an additional low-temperature contribution to the multi-phonon Raman response. This excess component can be parametrized using the same Arrhenius-type expression, yielding an activation energy of approximately 15~meV, consistent with the energy scale extracted from the single-phonon analysis.

The low-energy feature near 50~cm$^{-1}$ shows a pronounced asymmetric lineshape. Representative spectra at 6~K and 60~K are shown in Fig.~\ref{fig:fig3}c. The data can be well described by a superposition of a Lorentzian and a Fano profile, indicating strong electron-phonon coupling \cite{Klein1983}. While the overall spectral weight of this mode increases upon cooling, the asymmetry of the lineshape is reduced at the lowest temperatures, as reflected by a change in the fitted Fano parameter from $q=-2.6$ at 60~K to $q=-5.5$ at 6~K.

\subsection*{Discussion}

The temperature-dependent Raman results presented here reveal a pronounced evolution of the vibrational response of Cs$_2$SnI$_6$ upon cooling, extending beyond the behavior expected for an ideal cubic structure. Three experimental observations are particularly salient: (i) a strong enhancement of phonon spectral weight at low temperatures, (ii) the emergence of additional low-energy modes at cryogenic temperatures, and (iii) a pronounced increase of multi-phonon scattering under near-resonant excitation conditions. 
A key observation is that both cubic-symmetry-allowed and cubic-symmetry-forbidden phonon modes share a common thermally activated intensity scale of 15–20~meV. This indicates that their temperature evolution is governed by a unified underlying process rather than mode-specific effects. Importantly, no experimental evidence for a \textit{global} symmetry-lowering phase transition is found over the investigated temperature range, in agreement with diffraction studies reporting an average cubic structure down to cryogenic temperatures \cite{Maughan2016}. The Raman data therefore point to local, short-range symmetry reduction that does not develop into long-range structural order.
The activation of the additional phonon modes can be understood within a unified framework that is consistent with both the experimental Raman data and first-principles calculations. Density functional theory shows that a relaxed, symmetry-lowering structure derived from the cubic $Fm\bar{3}m$ phase naturally exhibits additional Raman-active modes at frequencies comparable to those observed experimentally (see Fig. 1). This demonstrates that relatively modest deviations from the idealized high-symmetry average structure are sufficient to lift Raman selection rules and enrich the low-energy vibrational spectrum. From the experimental perspective, this is consistent with the local sensitivity of Raman scattering, which probes instantaneous or short-range structural configurations even when the average crystal structure remains cubic.

Within this framework, one may envisage the following scenario. At elevated temperatures, the structure explores a broad manifold of locally distorted configurations \cite{Schilcher2021} that average to the high-symmetry cubic structure, resulting in weak or strongly broadened signatures of symmetry-forbidden modes. Upon cooling, dynamic averaging over locally distorted configurations is progressively reduced, leading to longer-lived short-range distortions. This enhances the sensitivity of Raman scattering to local symmetry breaking and increases the visibility of modes that are inactive in the ideal structure.
In this sense, the DFT-relaxed structure provides a useful static reference for the type of local symmetry breaking that can give rise to the observed additional phonon features, while the temperature dependence of the Raman intensities and lineshapes reflects the evolving contribution of these locally distorted configurations to the vibrational response.

The pronounced enhancement of multi-phonon scattering under near-resonant excitation further highlights the strong coupling between electronic excitations and atomic vibrations in this soft and anharmonic material. At elevated temperatures, the ratio of two-phonon to single-phonon scattering follows the expected Bose population behavior. At low temperatures, however, an additional contribution emerges that cannot be accounted for by population effects alone. Similar enhancements of higher-order Raman processes have been reported in other halide perovskites under resonant excitation and are commonly associated with strong electron–phonon coupling \cite{Iaru2021, Xu2022}. In Cs$_2$SnI$_6$, the coexistence of strong multi-phonon scattering, the activation of symmetry-forbidden modes, and the pronounced temperature dependence of phonon spectral weight points to a highly anharmonic structure that is strongly coupled to electronic degrees of freedom.


\subsection*{Conclusions} The temperature-dependent Raman measurements reveal a pronounced evolution of the vibrational response of Cs$_2$SnI$_6$ upon cooling that cannot be fully described within the framework of a typical cubic structure. In addition to a strong enhancement of phonon spectral weight at low temperatures, cubic-symmetry-forbidden modes emerge and multi-phonon scattering becomes significantly enhanced under near-resonant excitation. The activation of modes near 50~cm$^{-1}$ and 86~cm$^{-1}$, at energies corresponding to infrared-active phonons of the cubic phase, indicates a relaxation of Raman selection rules on local length scales. The absence of any global symmetry-lowering transition in diffraction studies implies that this symmetry reduction remains short-ranged and does not develop into long-range structural order.

First-principles calculations show that a relaxed, symmetry-lowering structure derived from the cubic phase naturally gives rise to additional Raman-active modes at comparable frequencies. The temperature dependence of both cubic-symmetry-allowed and cubic-symmetry-forbidden modes follows a common Arrhenius-type behavior with an energy scale of 15–20 meV, pointing to a shared thermally activated process governing the Raman visibility of locally distorted configurations. The simultaneous enhancement of multi-phonon scattering under near-resonant excitation further underscores the strong coupling between electronic excitations and lattice vibrations in this soft and anharmonic material. Together, these findings establish temperature-dependent Raman spectroscopy as a powerful probe of local symmetry breaking and electron–phonon interactions in vacancy-ordered halide perovskites.

\section{Acknowledgments}
\begin{acknowledgments}
The authors thank Nathan R. Wolf (Stanford University) for support with measurements of the experimental Cs$_2$SnI$_6$ structure at 100 K. Aspects of the data interpretation (N.J.W., M.F.T.) were supported by the Center for Hybrid Organic Inorganic Semiconductors for Energy, an Energy Frontier Research Center funded by the Office of Basic Energy Sciences, Office of Science within the U.S. Department of Energy (DOE).
J.A.V. acknowledges fellowship support from the Stanford University Office of the Vice Provost of Graduate Education and the National Science Foundation Graduate Research Fellowship Program under grant no. DGE-1656518.
T.G. and D.R. were supported by the U.S. Department of Energy, Office of Basic Energy Sciences, Office of Science, under Contract No. DE-SC0024117. Apparatus customization and  preliminary measurements conducted by C.S. were supported by the NSF grant No DMR-2210126.
X.Q., G.J.G., and V.B. were supported by the National Science Foundation under Award Number 2323803 for the first-principles simulations included in this work.
G.J.G. was also supported by the National Science Foundation under grant DGE-2022040 for computational data analysis and additional first-principles simulations included in this work.
X.Q. acknowledges the computational resources provided by the National Energy Research Scientific Computing Center (NERSC), a U.S. Department of Energy (DOE) Office of Science User Facility, operated under Contract No. DE-AC02-05CH11231.
G.J.G. acknowledges computational resources sponsored by the Department of Energy and located at the National Laboratory of the Rockies.
H.I.K. acknowledges funding through the DOE Office of Basic Energy Sciences, Division of Materials Science and Engineering under contract no. DE-AC02-76SF00515 (sample preparation).
\end{acknowledgments}

\section{Data Availability}
All computational structure optimizations, phonon calculations, and Raman intensity calculations can be obtained from the Duke Data Repository. The atomic geometries for the cubic $Fm\bar{3}m$ structure used for initial phonon calculations and the computationally relaxed, lower-symmetry structure used for phonon and Raman intensity calculations can be found at the end of the SI. These structures, along with experimental Raman data at 6K and 295K, can be found in the HybriD$^3$ database. Additional relevant data can be obtained from the corresponding author or the first author upon reasonable request.

\section{Declaration of Interests}
The authors declare no competing interests.

\bibliography{Literature}

\end{document}